\documentclass[preprint]{aastex61}
\usepackage{epstopdf}
\usepackage[toc]{appendix}
\usepackage{color}
\usepackage{lineno}
\received{\today}
\submitjournal{ApJ}

\shorttitle{Universality of current sheet thinning}
\shortauthors{Artemyev et al.}

\begin{document}

\title{Thin current sheet formation: comparison between Earth's magnetotail and coronal streamers}  

\correspondingauthor{Anton V. Artemyev}
\email{aartemyev@igpp.ucla.edu}

\author{Artemyev A. V.}
\affiliation{Department of Earth, Planetary, and Space Sciences and Institute of Geophysics and Planetary Physics, \\University of California, Los Angeles, California, USA}

\author{Réville V.}
\affiliation{IRAP, Université Toulouse III—Paul Sabatier, CNRS, CNES, Toulouse, France}

\author{Ivan Zimovets}
\affiliation{Space Research Institute, RAS, Moscow, Russia}


\author{Nishimura Y.}
\affiliation{Department of Electrical and Computer Engineering and Center for Space Sciences, Boston University, Boston, MA USA}

\author{Velli M.}
\affiliation{Department of Earth, Planetary, and Space Sciences and Institute of Geophysics and Planetary Physics, \\University of California, Los Angeles, California, USA}


\author{Runov A.}
\affiliation{Department of Earth, Planetary, and Space Sciences and Institute of Geophysics and Planetary Physics, \\University of California, Los Angeles, California, USA}

\author{Angelopoulos V.}
\affiliation{Department of Earth, Planetary, and Space Sciences and Institute of Geophysics and Planetary Physics, \\University of California, Los Angeles, California, USA}


\begin{abstract}
Magnetic field line reconnection is a universal plasma process responsible for the magnetic field topology change and magnetic field energy dissipation into charged particle heating and acceleration. In many systems, the conditions leading to the magnetic reconnection are determined by the pre-reconnection configuration of a thin layer with intense currents -- otherwise known as the thin current sheet. In this study we investigate two such systems: Earth's magnetotail and helmet streamers in the solar corona. The pre-reconnection current sheet evolution has been intensely studied in  the magnetotail, where in-situ spacecraft observations are available; but helmet streamer current sheets studies are fewer, due to lack of in-situ observations -- they are mostly investigated with numerical simulations  and information that can be surmised from remote sensing instrumentation. Both systems exhibit qualitatively the same behavior, despite their largely different Mach numbers,  much higher at the solar corona than at the  magnetotail. Comparison of spacecraft data (from the magnetotail) with numerical simulations (for helmet streamers) shows that the pre-reconnection current sheet thinning, for both cases,  is primarily controlled by plasma pressure gradients. Scaling laws of the current density, magnetic field, and pressure gradients are the same for both systems. We discuss how magnetotail observations and kinetic simulations can be utilized to improve our understanding and modeling of the helmet streamer current sheets.
\end{abstract}


\section{Introduction}
Magnetic field-line reconnection is a key plasma process responsible for energy transformation and charged particle acceleration in solar corona \citep[see, e.g.,][]{Priest&Forbes02,Aschwanden02,Zharkova11SSR} and planetary magnetospheres \citep[see, e.g., reviews by][]{Jackman14,book:Gonzalez&Parker,bookBirn&Priest07}. Spatially localized regions of intense plasma currents, i.e., current sheets, are believed to be the primary regions where magnetic reconnection takes place  \citep{bookBiskamp00,bookParker94,bookBirn&Priest07}. The spatial spatial structure (topology) of currents sheets in the solar corona \citep{Priest85,bookPriest16} is generally much more complex than that in planetary magnetospheres. However, one notable exception is the helmet streamer current sheet, which shares many similarities in configuration with the most-investigated magnetotail current sheet \citep[see discussion][]{Syrovatskii81,Terasawa00:AdSR, Reeves08:solar}.  

A zero order approximation of the current sheet configuration in both the magnetotail and helmet streamers is provided by the Grad-Shafranov equation $\nabla\times\nabla \times A=4\pi j$, in which  a single vector potential component $A(x,z)$ determines the  field and the current density $j$ in  the systems' 2D plane geometry (for the solar corona $x$ refers to the  radial distance $r$, and $z$ is along the direction of solar colatitude). However, the two systems differ in that for the magnetotail $j=-dP/dA$ is due to diamagnetic drifts of the hot plasma, that which contributes to pressure $P$ \citep{Schindler&Birn78, Birn77}, whereas for the helmet streamer $j$ is altered by the gravitational force $F_G(r)$ and dynamical pressure gradients $\sim {\bf v}\nabla {\bf v}$ \citep{Pneuman&Kopp71}. Although such flows can be included into a generalized magnetotail current sheet equilibrium with incompressible plasma \citep{Birn91:pop,Birn92}, in the absence of conductivity such a generalization is possible only for field-aligned ${\bf v}\propto{\bf B}\sim-\nabla A$ flows \citep[see details in][]{Nickeler&Wiegelmann10}. Such  currents (produced by gradients of thermal and dynamical pressures) stretch magnetic field lines away from the Earth/Sun and reduce the magnetic field fall-off with radial distance \citep[e.g.,][]{Washimi87,Wang&Bhattacharjee99}. A common methodology for reconstructing such current sheet configurations is to first identify system symmetries corresponding to integrals \citep[e.g.,][]{Yeh84,Birn92,Hodgson&Neukirch15}, which enables the  derivations of the generalized  Grad-Shafranov equation \citep[see examples in][]{Neukirch97,Neukirch95,Cheng92,Zaharia05}.

One additional important plasma property that needs to be incorporated in the equations describing helmet streamer current sheets, but is not typically considered  in similar descriptions of the magnetotail, is temperature inhomogeneity due to the plasma thermal conductivity \citep{Yeh&Pneuman77,Cuperman90}. The interplay of plasma temperature gradients and dynamical pressure gradients at different plasma beta, $\beta=8\pi P/B^2$, significantly affects the evolution  of helmet streamer current sheets \citep[see][]{Steinolfson82}. \citet{Cuperman92,Cuperman95} showed a significant difference of current sheet configurations modeled with polytropic pressure (typical equation of state for the magnetotail current sheet models) and with the pressure determined by the thermal conductance (typical equation of state for the Solar corona current sheet models).

The most advanced fluid models of magnetotail and helmet streamer current sheets describe 3D magnetic field configurations \citep[e.g.,][]{Cuperman93, Linker01, Birn04MHD} or include a multi-fluid Hall physics \citep[e.g.,][]{Endeve04,Ofman11,Ofman15,Rastaetter99,Yin&Winske02}. However, kinetic physics, whereby the particle distribution functions and not just their moments are included in the dynamical evolution of the system, cannot be incorporated in system-scale models. Such physics is mostly investigated and validated in small-scale (spatially localized) simulations available and justified for the magnetotail \citep[see][]{Pritchett&Coroniti94,Pritchett&Coroniti95,Birn&Hesse05,Birn&Hesse14,Liu14:CS}. 

Both the magnetotail and helmet streamer current sheets are primary regions of magnetic field-line reconnection that is driven externally in both systems \citep[e.g.,][]{Guo96, Guo&Wu98,Verneta94, Airapetian11, Birn96, Birn04MHD}. Thus, the construction of accurate current sheet configurations is important to set realistic initial conditions for dynamical simulations of the magnetic reconnection and plasma heating \citep[see discussion in][]{Wu&Guo97,Wu&Guo97:AGU}. A slow current sheet evolution, i.e., the current sheet thinning \citep[e.g.,][]{Wiegelmann&Schindler95,Birn98:cs}, plays a crucial role in determining the pre-reconnection current sheet configuration, which further controls the efficiency of the magnetic field energy dissipation and plasma heating. There is a good similarity in theoretical approaches for construction of such current sheet configurations \citep[see discussion in][]{Schindler83,Neukirch97} and investigation of the current sheet stability \citep[see discussion in][]{Birn&Hesse09}. An open question here is whether plasma flows play more important roles for the stability of the coronal current sheets \citep[see, e.g,][]{Dahlburg&Karpen95,Einaudi99,Lapenta&Knoll03,Feng13} than for the magnetotail current sheet stability \citep[see discussion in][]{Shi21:jgr:tearing}. In hot magnetotail plasma, such flows may drive a slow current sheet evolution \citep[see discussion in][]{Nishimura&Lyons16:flows, Pritchett&Lu18}, but cannot determine the current sheet configuration, whereas in the corona plasma flows may contribute to the current sheet configuration \citep[see discussion in][]{Pneuman&Kopp71, Neukirch95:pop}.

\begin{figure}
\centering
\includegraphics[width=0.6\textwidth]{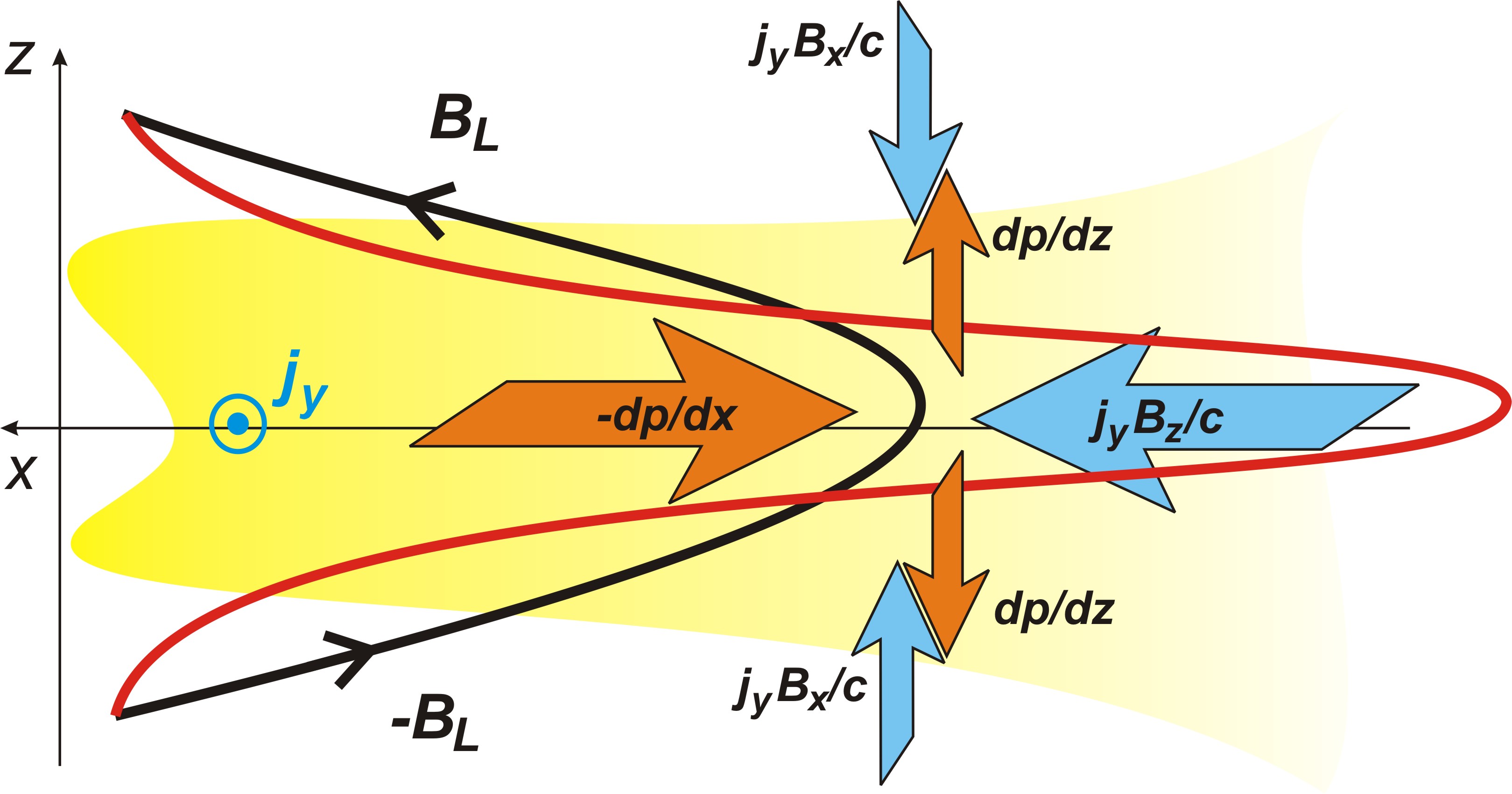}
\caption{\label{magnetotail:schematic} Schematic view of the magnetotail current sheet thinning with the main physical characteristics indicated. Two magnetic field lines are shown: parabolic black line before thinning and stretched red line after thinning.}
\end{figure}

Magnetic field configuration of helmet streamer current sheets are observed in-situ only at large distances from the current sheet formation region \citep[see, e.g., discussion in][]{Pneuman72,Gosling81, Eselevich&Filippov88,Bavassano97}, and mostly probed remotely \citep[e.g.,][]{Woo97,Gopalswamy03}. However, reconnection-related plasma injections from helmet streamer current sheets can be well captured \citep[see most recent results in][and references therein]{Korreck20,NievesChinchilla20,Lee21:streamers,Liewer21}, which underlines the importance of investigating the reconnection and pre-reconnection dynamics of helmet streamer current sheets. In contrast to helmet streamer current sheets, the configuration of the magnetotail current sheet is well studied with multiple in-situ plasma and magnetic field measurements \citep[see reviews by][and references therein]{Baumjohann07,Artemyev&Zelenyi13,Petrukovich15:ssr,Sitnov19}. Therefore, observational aspects of the magnetotail current sheet thinning (formation of the thin current sheet before reconnection onset, see \citet{Runov21:jastp} and references therein) can be compared with models of helmet streamer current sheets. 

In this study we focus on such a comparison of magnetotail current sheet observations and helmet streamer current sheet simulations, especially during the current sheet thinning. In the magnetotail current sheet configuration, such slow thinning can be treated as a 2D process with a fine balance between the magnetic field line tension and plasma pressure gradient (see schematic \ref{magnetotail:schematic}). To simulate the same current sheet thinning in the helmet streamer, we use a fluid model \citep[see details in][]{Reville20:ApJ,Reville20:ApJS} based on PLUTO code \citep{Mignone07}. We compare temporal profiles of stress balance components, magnetic field and current density magnitudes for the two systems and show similarities of the thin current sheet formation.

The paper is structured as follows: Sections~\ref{sec:magnetotail} and \ref{sec:simulations} describe the basic spacecraft datasets for the magnetotail current sheet and the simulation setup for the helmet streamer; Section~\ref{sec:balance} compares stress balance dynamics in current sheet thinning for two systems, and Section~\ref{sec:conclusions} discusses main results.

\section{Spacecraft observations and simulation datasets}
\label{sec:data}

\subsection{Magnetotail observations}\label{sec:magnetotail}
Spacecraft observations of the magnetotail current sheet thinning before the reconnection onset form the basement for the magnetosphere substorm concept \citep[see][and references therein]{Baker96,Angelopoulos08}. One of the key elements of such thinning is the current density $j_y$ increase (see geometrical schematic in Fig. \ref{magnetotail:schematic}), which can be only reliably measured by multi-spacecraft with the curlometer technique \citep{Dunlop02,Runov05:pss}. Therefore, statistical investigations of the current sheet thinning start with the Cluster mission \citep{Escoubet01} having four spacecraft that can probe magnetic fields and magnetic field gradients (currents and current sheet thickness) in the thinning current sheet  \citep[see][]{Kivelson05,Petrukovich07,Petrukovich13,Snekvik12}. However, due to their polar orbit, Cluster spacecraft did not spend long intervals in the slowly thinning current sheet, which are better probed by the THEMIS mission \citep{Angelopoulos08:ssr} on equatorial orbits. Three THEMIS spacecraft may be in the same azimuthal $y$ plane and locate above ($B_x>0$), below ($B_x<0$), and around ($B_x\sim 0$) the current sheet neutral plane, the plane of the main magnetic field ($B_x$) reversal. Thus, gradients $\partial B_x/\partial z$, $\partial B_z/\partial z$ contributing to the current density $j_y$ (see Fig. \ref{magnetotail:schematic}) can be calculated from the differences of magnetic field at three spacecraft \citep[see][]{Artemyev16:jgr:thinning}. Magnetic field $B_x$ is a good proxy of the spacecraft distance relative to the neutral plane (where $B_x=0$) and plasmasphere lobes (where $B_x$ reaches maxima of $B_L$). In the magnetotail current sheet configuration $\partial B_x/\partial z\gg \partial B_z/\partial z$, and the lobe field $B_L$ can be estimated from the pressure balance $\langle{B_x^2+8\pi P}\rangle=B_L^2$, where $P$ is the plasma (ion and electron) pressure and $\langle…\rangle$ is the average over the interval of the current sheet crossing \citep[see discussion of how $B_L$ is reliable in][]{Petrukovich99}. 

\begin{figure}
\centering
\includegraphics[width=1\textwidth]{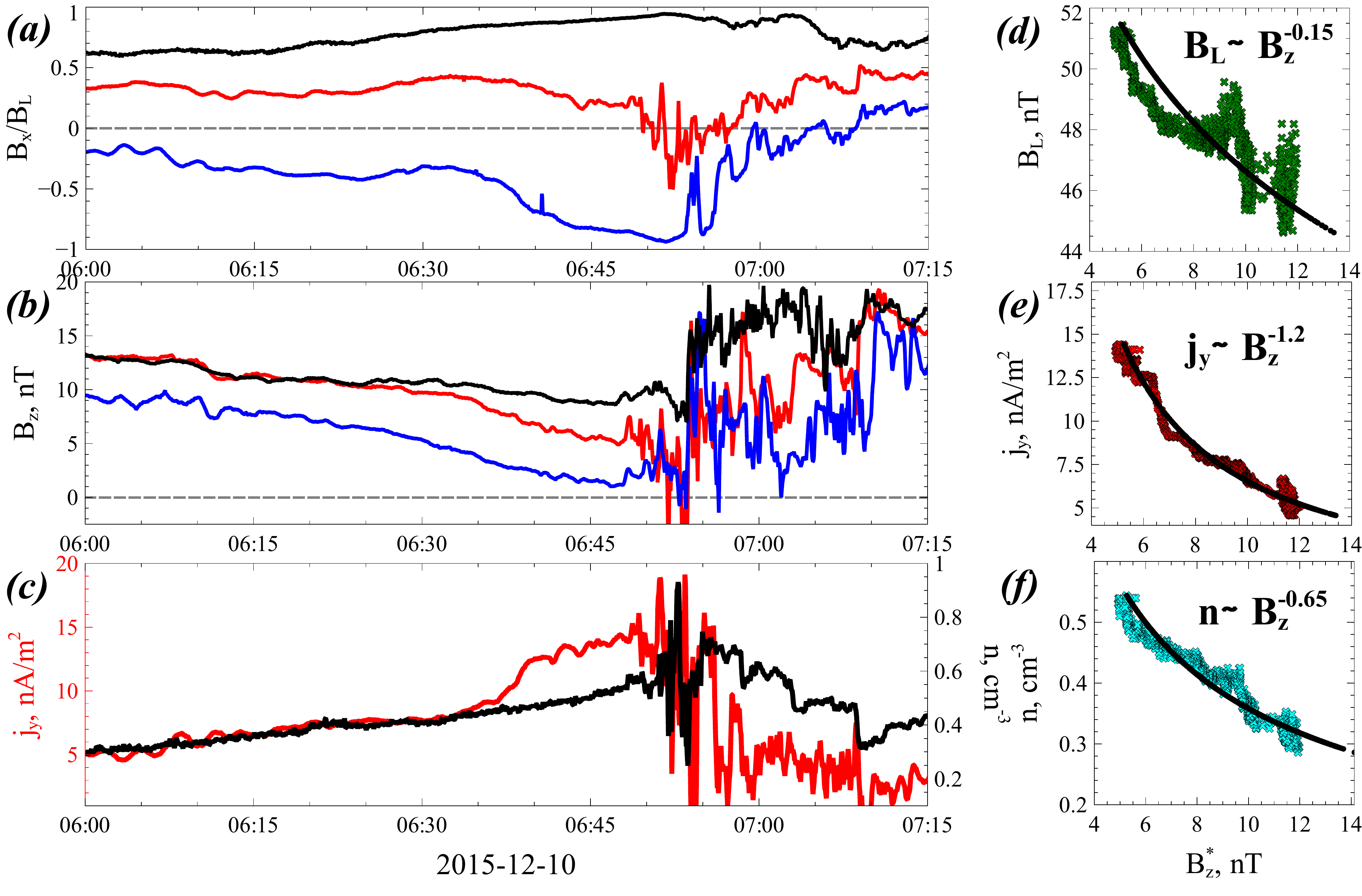}
\caption{\label{magnetotail:01} Example observations of the magnetotail current sheet thinning (before 06:50 UT) and thin current sheet destruction by plasma flows from the reconnection site (at 06:50 UT). Panels (a) and (b) show $B_x$ and $B_z$ magnetic field components measured by THEMIS A (black), D (red), and E (blue). Magnetic field $B_x$ is normalized to the lobe magnetic field $B_L=\max |B_x|$ estimated from the pressure balance and plasma pressure measurements. Panel (c) shows current density $j_y=(c/4\pi)\cdot\left(\partial B_x/\partial z-\partial B_z/\partial x \right)$ and plasma density (measured by THEMIS D). Panels (d,e,f) show $B_L$, $j_y$, and $n$ as a function of $B_z^*$ ($B_z^*$ is the $B_z$ measured by the THEMIS spacecraft closest to the neutral plane) during the current sheet thinning. We also show fits of $B_L(B_z^*)$, $j_y(B_z^*)$, $n(B_z^*)$ by power law functions.}
\end{figure}

Figure \ref{magnetotail:01} (a) shows an example THEMIS observation of magnetic fields in the thinning current sheet. Magnetic field is measured by THEMIS fluxgate magnetometer \citep{Auster08:THEMIS} and normalized to $B_L$, which is calculated with plasma measurements from electrostatic analyzer \citep[for energies below 30keV, see][]{McFadden08:THEMIS} and solid state detector \citep[for energies between 30 and 500 keV, see][]{Angelopoulos08:sst}. THEMIS D (red) is located close to the neutral sheet and observe $B_x/B_L$ oscillations around zero, whereas THEMIS E (blue) and A (black) are located below and above the neutral plane and observe $|B_x/B_L|$ increases (the typical signature of the current sheet thinning) and $j_y\sim \Delta B_x$ growth. Simultaneous with $B_x/B_L$ variations, all three THEMIS probes observe $B_z$ decreases (see panel (b)), i.e., the current sheet thinning is accompanied by magnetic field line stretching \citep[see discussion in][]{Petrukovich07}. Thinning ends at around 06:50 UT by the reconnection onset downtail (at large radial distances). Plasma flows propagating from the reconnection site bring strong $B_z$ increases (so-called dipolarization fronts \citep[see][]{Nakamura02,Nakamura09,Runov09grl,Sitnov09} that lead to $B_x/B_z$ ratios close to the nominal dipole, $B_x/B_z\sim 1$) and destroy the thin current sheet (see $|B_x/B_L|$ decreases at THEMIS E and A).

The current sheet thinning and following destruction of thin current sheet can also be seen in the temporal profile of the current density $j_y$ shown in panel (c). Before 06:50 UT, $j_y$ goes up, but then quickly drops to zero after the thin current sheet is destroyed by the dipolarization front. The current density increase is accompanied by plasma density increase, i.e., thinning current sheet becomes denser (see panel (c)). Statistical observations of the magnetotail current sheet thinning show that this density increase is due to the arrival of new cold plasma populations from deep magnetotail or ionosphere \citep[e.g.,][]{Artemyev19:jgr:globalview}, and such a density increase is associated with a plasma temperature decrease \citep[see][]{Artemyev16:jgr:thinning,Yushkov21}. Plasma cooling in the thinning current sheet contradicts to the concept of the compressional thinning \citep[see][]{Schindler&Birn86,Schindler&Birn99}, and indeed the current density increase $j_y$ is much stronger than $B_L$ increase in thinning current sheets \citep[see][]{Artemyev16:jgr:thinning,Sun17:cs_pressure}. Thus, the formation of thin current sheet may mostly involve the interval reconfiguration of the current sheet with only a weak external driver. 

As the equatorial magnetic field $B_z$ monotonically decreases during the current sheet thinning, we can use it, instead of time, to trace the evolution of the current sheet configuration. Figure \ref{magnetotail:01} (d,e,f) shows $B_L(B_z^*)$, $j_y(B_z^*)$, and $n(B_z^*)$ profiles for the interval with $B_z^*$ decrease ($B_z^*$ is the $B_z$ measured by the THEMIS spacecraft closest to the neutral plane). The current density magnitude grows almost linearly with $1/B_z^*$; this means that the magnetic field line tension force $j_yB_z$, which balances the plasma pressure gradient at the equator $\partial P/\partial x$, does not vary much during the current sheet thinning. The lobe magnetic field (or equatorial plasma pressure $\approx B_L^2/8\pi$) varies with $B_z$ slowly, and thus we indeed deal with the current sheet thinning: the current sheet thickness $L\approx cB_L/4\pi j_y \propto B_z$ goes down. The plasma density increases much faster then $B_L^2$, and this confirms the plasma temperature decrease $T=P/n=B_L^2/8\pi n\propto B_z^{1/3}$.

\begin{figure}
\centering
\includegraphics[width=1\textwidth]{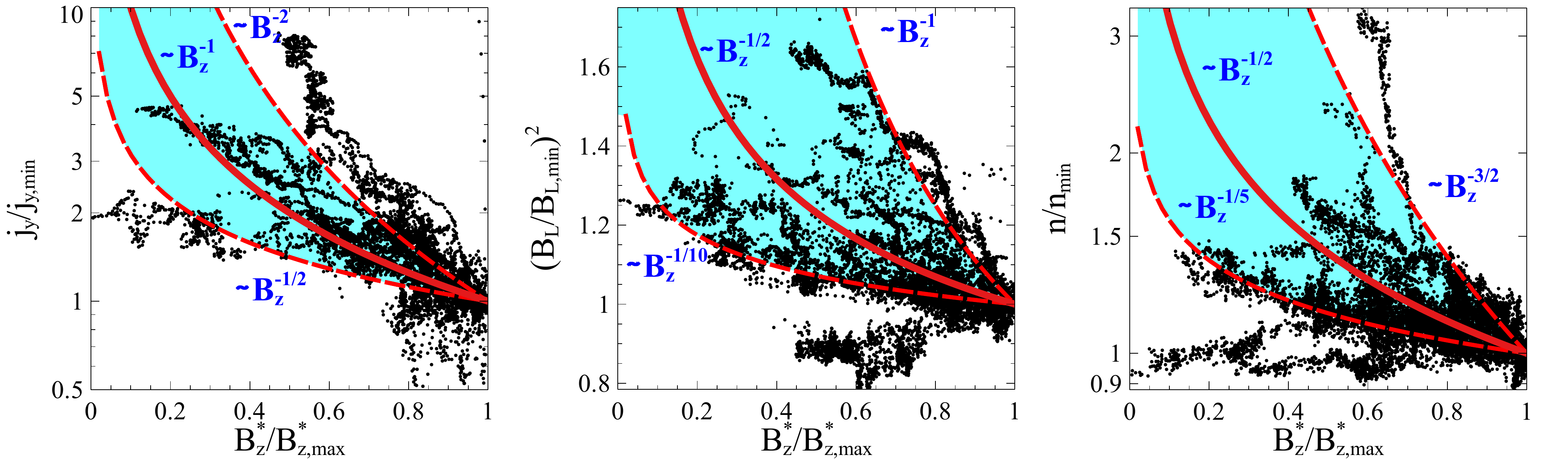}
\caption{\label{magnetotail:02} Statistical characteristics of current density, plasma density, and magnetic field magnitude evolution in thinning current sheets ($B_z$ decreases with time). For each event from \citep{Artemyev16:jgr:thinning}, we normalize the current density, lobe field, and plasma density to their nominally minimum values (the initial moment of time when spacecraft enter the thinning current sheet).}
\end{figure}

Although the $B_z$ decrease and $j_y$ increase are characteristic features of the magnetotail current sheet thinning, rates of these processes may vary from event to event. Figure \ref{magnetotail:02} shows statistical THEMIS observations of the current sheet thinning \citep[for the database details ee]{Artemyev16:jgr:thinning}. The current density variation is generally slightly faster than $\propto  1/B_z$, whereas the lobe field $B_L$ increases generally slower than $\propto  1/B_z^{1/2}$, and this difference means that the $j_y$ increase is accompanied by the current density decrease. Moreover, for many events  $B_L$ does not change ($B_L\propto  1/B_z^{1/4}$), indicating that the thin current sheet with intense $j_y$ is growing inside a thicker plasma sheet supporting $B_L$, i.e., we deal with the so-called embedded current sheets \citep[see][]{Runov06,Artemyev10:jgr,Petrukovich09}. The typical thickness of an embedded thin current sheet is the proton gyroradius $\rho_p=\sqrt{m_pc^2T_p}/ eB_0$, as determined by the field $B_0$ at the boundary of the thin current sheet, $L\approx cB_0/4\pi j_y \approx \rho_p$ \citep{Artemyev11:jgr,Petrukovich15:ssr}. Although for most intense current sheets $B_0$ may reach $B_L$, i.e., the thin current sheet contains the entire cross-sheet current density, it is more usual to have $B_0\sim B_L/3$ \citep{Artemyev10:jgr}. 

The plasma density increase is quite typical for the current sheet thinning (see left panel in Fig. \ref{magnetotail:02}), but the rate of this increase is smaller than the $j_y$ increase rate. Thus, the total bulk speed of the current carriers (difference between ion and electron drifts) is growing, as $j_y/en\propto 1/B_z^{1/2}$, in the thinning current sheet. 

In this work, we will compare observational trends of $j_y$, $n$, and plasma pressure $B_L^2/8\pi$ evolutions with the results of fluid simulations of the current sheet thinning in the helmet streamer configuration. The main difference of these two systems, magnetotail and helmet streamer, is the strong plasma flows absent in the pre-reconnection magnetotail, but potentially contributing to the helmet streamer current sheet configuration.

\subsection{Streamers' simulations}\label{sec:simulations}
To simulate the helmet streamer current sheet, we use a 2.5D MHD code that includes effects of wave pressure and turbulent heating \citep[see details in][]{Reville20:ApJ,Reville20:ApJS} based on the PLUTO code \citep{Mignone07}. The simulation shown here repeats one from \citet{Reville20:ApJ}, with Lundquist number $S=10^5$.  The simulation is initialized with a dipole field, while mechanisms of coronal heating and wind acceleration act to thin the heliospheric current sheet located at $\theta=\pi/2$. After about $90h$ of simulation time, a tearing instability is triggered in the current sheet. In the following, we focus on the thinning process. 

\begin{figure}
\centering
\includegraphics[width=0.9\textwidth]{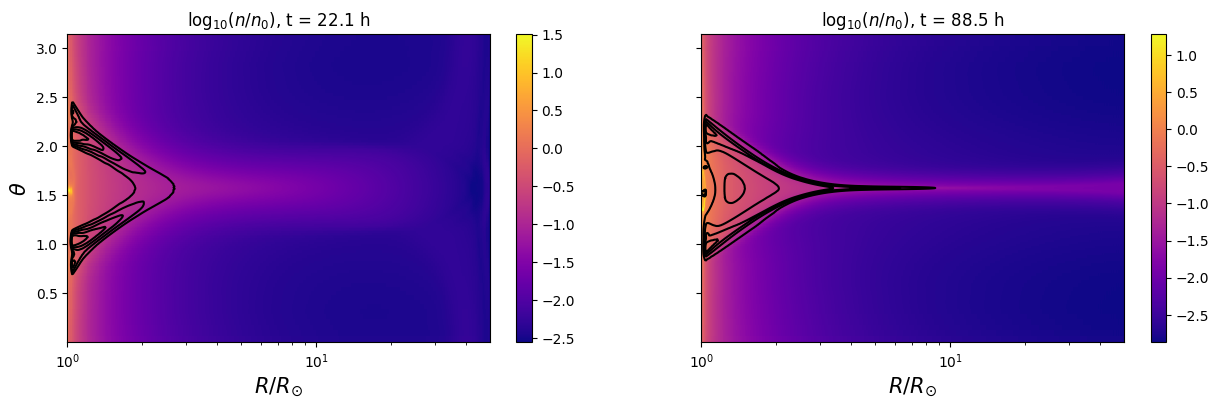}
\caption{\label{streamer:01} 2D view of the thin current sheet formation in helmet streamer geometry. For two moments of time we show $j_\varphi=const$ contours (black) and plasma density (color). The first moment (left) is the initial state and the final moment (right) is the current sheet right before the reconnection onset. The normalization factor for the plasma density is $n_0=10^8$cm$^{-3}$.}
\end{figure}

Figure \ref{streamer:01} shows the plasma density distribution and current density contours in the $(r,\theta)$ plane for the initial and final moment of the simulation, right before the reconnection onset at $r\sim 5r_\odot$. The radial direction $r$ is an analog of $x$ coordinate in the magnetotail current sheet, and $r\cos\theta$ is an analog of $z$ coordinate (here $\theta$ is colatitude). A thin current sheet clearly develops in the stretched magnetic field lines along the streamer, and we aim to compare dynamic properties of this current sheet thinning with the magnetotail observational dataset. To simplify the comparison we will focus on the stress balance equations along the radial direction in the neutral plane $\theta=\pi/2$:
\begin{equation}
\rho\frac{{\partial v_r }}{{\partial t}} + \underbrace {\rho v_r \frac{{\partial v_r }}{{\partial r}} + \rho \frac{v_\theta }{r}\frac{{\partial v_r }}{{\partial \theta }} - \frac{{ v_\theta ^2 }}{r}}_{n\left[ {\left( {{\bf v}\nabla } \right){\bf v}} \right]_r } + \underbrace {\frac{{\partial P}}{{\partial r}} + \frac{1}{{8\pi }}\frac{1}{{r^2 }}\frac{{\partial r^2 B_\theta ^2 }}{{\partial r}}}_{\nabla \left( {P + B^2 /8\pi } \right)_r } = \underbrace { - \frac{{g\rho }}{{r^2 }} + \frac{{B_\theta  }}{{4\pi r}}\frac{{\partial B_r }}{{\partial \theta }}}_{\left[ {\left( {{\bf B}\nabla } \right){\bf B}} \right]_r /4\pi  - g\rho /r^2 } \label{eq:streamer}
\end{equation}
where $v_r$ and $ v_\theta $ are radial and azimuthal velocity components, $g$ is the gravitational constant of the Sun, and $\rho=nm_p$. Equation (\ref{eq:streamer}) shows that for slow (adiabatic) current sheet thinning, three main stress balance terms should be in balance: gradients of dynamic pressure $n\left[({\bf v}\nabla){\bf v}\right]_r$, total pressure $\nabla (P+B^2/8\pi)$, and magnetic field line tension force $\left[({\bf B}\nabla){\bf B}\right]$ with the gravitation correction $-gn/r^2$. The analog of Eq. (\ref{eq:streamer}) in the magnetotail current sheet configuration can be written as
\begin{equation}
\rho \frac{{\partial v_x }}{{\partial t}} + \underbrace {\rho \left( {v_x \frac{{\partial v_x }}{{\partial x}} + v_z \frac{{\partial v_x }}{{\partial z}}} \right)}_{ \rho \left[ {\left( {{\bf v}\nabla } \right){\bf v}} \right]_x } + \underbrace {\frac{{\partial P}}{{\partial x}} + \frac{1}{{8\pi }}\frac{{\partial B_z^2 }}{{\partial x}}}_{\nabla \left( {P + B^2 /8\pi } \right)_x  \approx \nabla _x P} = \underbrace {\frac{{B_z }}{{4\pi }}\frac{{\partial B_x }}{{\partial z}}}_{\left[ {\left( {{\bf B}\nabla } \right){\bf B}} \right]_x /4\pi  \approx j_y B_z /c} \label{eq:tail}
\end{equation}
In the hot magnetotail plasma $P\gg \rho v^2/2$, there is no contribution of plasma flows to the current sheet configurations before the reconnection onset \citep{Rich72}. Note that $\nabla P\gg\rho({\bf v}\nabla){\bf v}$ does not work for complex quasi-1D current sheet configurations that include nongyrotropic pressure terms \citep[see discussion in][and reference therein]{Steinhauer08,Artemyev21:grl}.

\section{Discussion of the force balance in thinning current sheets}\label{sec:balance}
\begin{figure}
\centering
\includegraphics[width=1\textwidth]{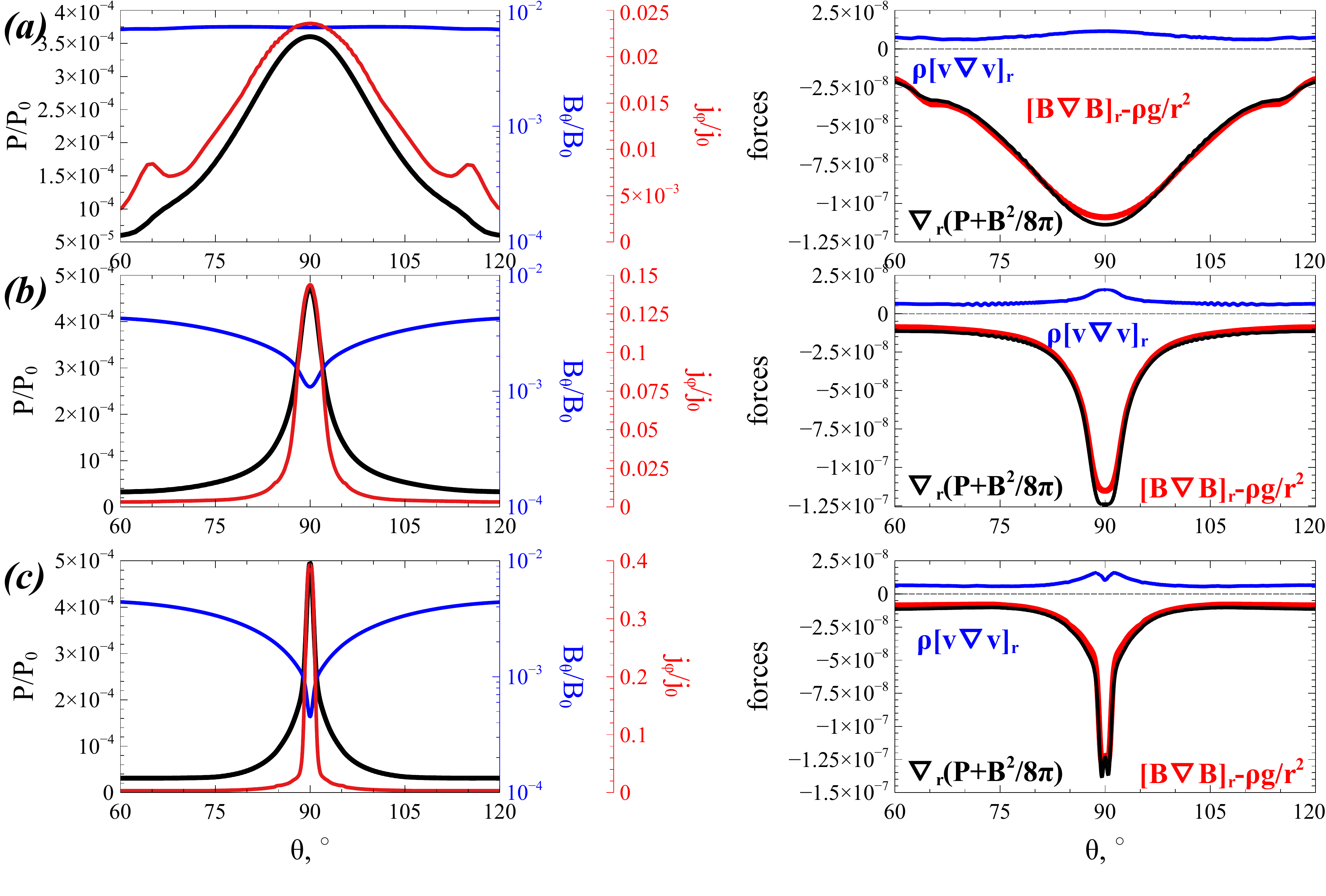}
\caption{\label{streamer:02} Left panels show cross-sheet $\theta$-profiles of current density $j_\varphi=\left(r^{-1}\partial B_r/\partial \theta-\partial B_\theta/\partial r \right)$, normal magnetic field $B_\theta$, and plasma pressure for three moments of time ($t=0$, $t=5$h, and $t=18$h). Right panels show cross-sheet $\theta$-profiles of the three main terms of the radial stress balance. All profiles are plotted at the radial distance of the magnetic reconnection region, which occurs right after $t=18$h. Magnetic field components are normalized on $B_0\approx 2$ Gauss, plasma pressure is normalized to $P_0=0.319$ dyne/cm$^2$, and the current density is normalized on $j_0=cB_0/4\pi r_\odot$. All stress balance terms (forces) are in units of 
kg/s$^2$/cm/$r_\odot$.}
\end{figure}

Figure \ref{streamer:02} shows cross-sheet profiles (along $\theta$) of main current sheet characteristics (magnetic field $B_\theta$, plasma pressure $P$, and current density $j_\varphi$) and main terms of the stress balance (dynamical pressure, total pressure, and magnetic field line tension force) at different times of the simulation. Initially, $\theta$ profiles show the single scale, and the stress balance is provided by $\nabla (P+B^2/8\pi)\approx \left[({\bf B}\nabla){\bf B}\right]/4\pi$. Note that profiles in Fig. \ref{streamer:02} are plotted at the radial distance where the magnetic reconnection will occur after $\sim 18$h of simulation time. 

The current sheet thinning is driven by a slight imbalance between the pressure gradient and the magnetic tension at the equator. As time goes on, a thin embedded current sheet develops in the central region of the simulation domain (see panel (b)). All three profiles of $B_\theta$, $P$, and $j_\varphi$ exhibit a gradient: strong peaks of $j_\varphi$ and $P$ (and minimum of $B_\theta$) around the center region ($\theta=\pi/2$), with almost unchanged levels (from the initial profile) elsewhere. The terms in the force balance equation also exhibit significantly different $\theta$-profiles, but the current sheet is still balanced as $\nabla (P+B^2/8\pi)\approx \left[({\bf B}\nabla){\bf B}\right]/4\pi$, with secondary (small-scale) peaks of $(P+B^2/8\pi)$, $\left[({\bf B}\nabla){\bf B}\right]/4\pi$ around $\theta=\pi/2$.

Right before the reconnection onset (see panel (c)), the $B_\theta$ minimum at $\theta=\pi/2$ almost reaches zero, whereas the peaks in current density and plasma pressure become even stronger. The pressure peak is embedded in the initial large-scale profile. Similar peaks are formed in the force balance components, with the same balance provided by $\nabla (P+B^2/8\pi)\approx \left[({\bf B}\nabla){\bf B}\right]/4\pi$. There is a small minimum of $\left[({\bf B}\nabla){\bf B}\right]/4\pi$ at the center region ($\theta=\pi/2$), which is likely balanced by the dynamic pressure gradient in the same region. However, magnitude of the dynamic pressure contribution to the stress balance remains small. 

\begin{figure}
\centering
\includegraphics[width=0.6\textwidth]{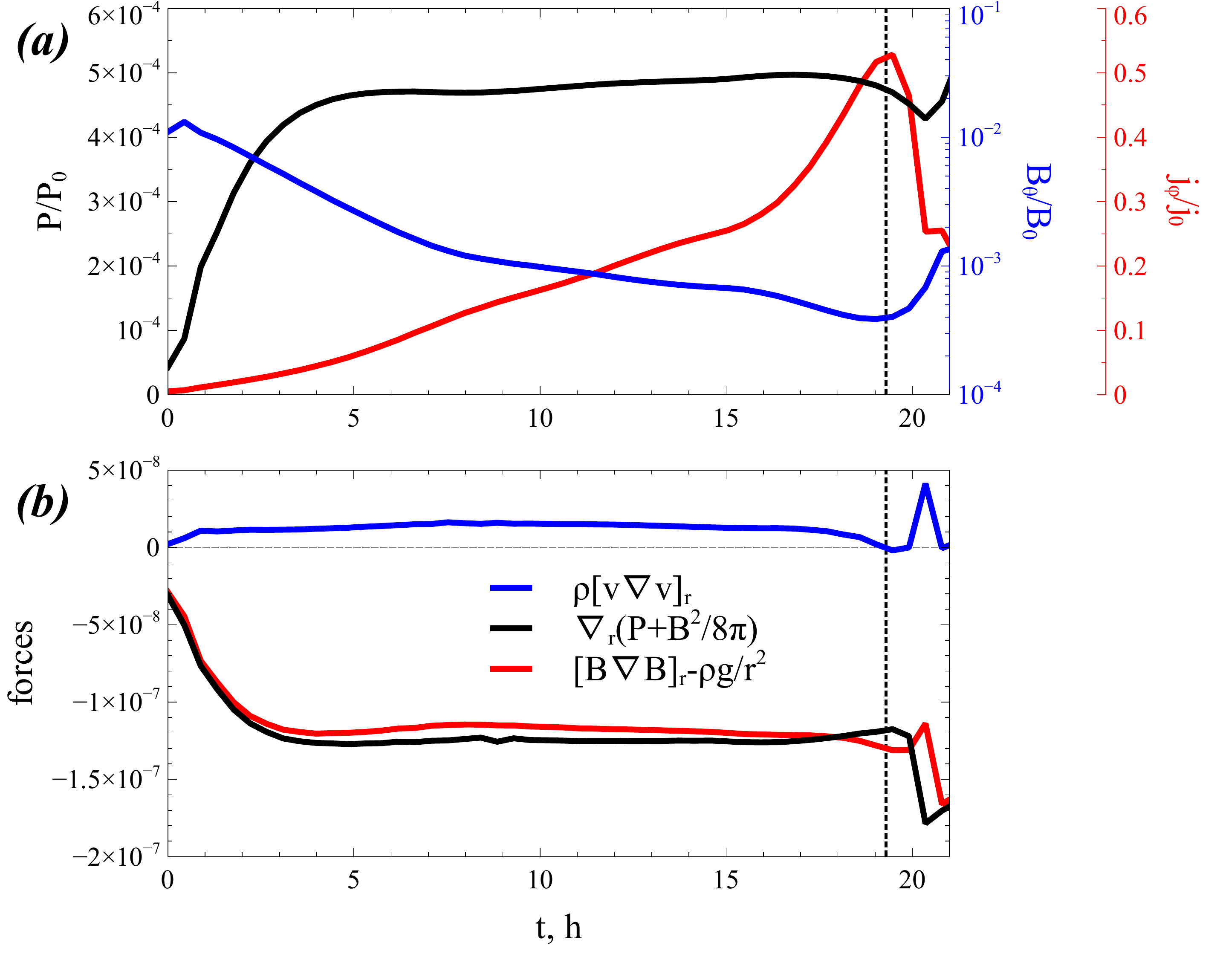}
\caption{\label{streamer:03} Evolution of the main current sheet characteristics and stress balance terms in the current sheet center, $\theta=\pi/2$. See details of the normalizations and units in the caption of Fig. \ref{streamer:02}.}
\end{figure}

Figure \ref{streamer:03} shows the evolution of the main current sheet characteristics and stress balance terms at the current sheet center, $\theta=\pi/2$. There is a clear increase of the current density $j_\varphi$ due to the formation of an embedded current sheet, which is accompanied by a decrease in $B_\theta$. The balance between $j_\varphi$ and $B_\theta$ variations results in an almost constant $j_\varphi B_\theta\approx \left[({\bf B}\nabla){\bf B}\right]_r$, which balances $\nabla (P+B^2/8\pi)_r$. Note that $\nabla (P+B^2/8\pi)_r$ does not vary because $P$ only increases weakly. Therefore, the current sheet thinning also means the current sheet stretching with $L_r\approx P/(\nabla P)_r$  increasing proportionally to $P$. Note for the same magnetic field configuration, \citet{Reville22:A&A} shows that the current sheet thinning is related to a pressure-driven instability, akin to the ballooning mode, whose timescales depend on the balance between magnetic curvature vector, gravity, and pressure gradients. Due to the stress balance along $\theta$, the current sheet thickness can be estimated as $L\approx cB_0/j_\varphi4\pi$, with $B_0\approx \sqrt{8\pi P}$. Thus, for the scale ratio we have $L_r/L \propto P j_\varphi/B_0 \propto B_0/B_\theta $, where $j_\varphi B_\theta \sim const$. As $P\approx B_0^2/8\pi$ increases weakly and  $B_\theta$ decreases, $L_r/L$ increases during the current sheet thinning, i.e., magnetic field lines stretching makes the current sheet more 1D (see also Fig. \ref{streamer:01}). This resembles the evolution of the magnetotail current sheet, where $L_x/L$ also goes down during the current sheet thinning \citep[see discussion in][]{Petrukovich13,Artemyev16:jgr:thinning}.

\begin{figure}
\centering
\includegraphics[width=0.6\textwidth]{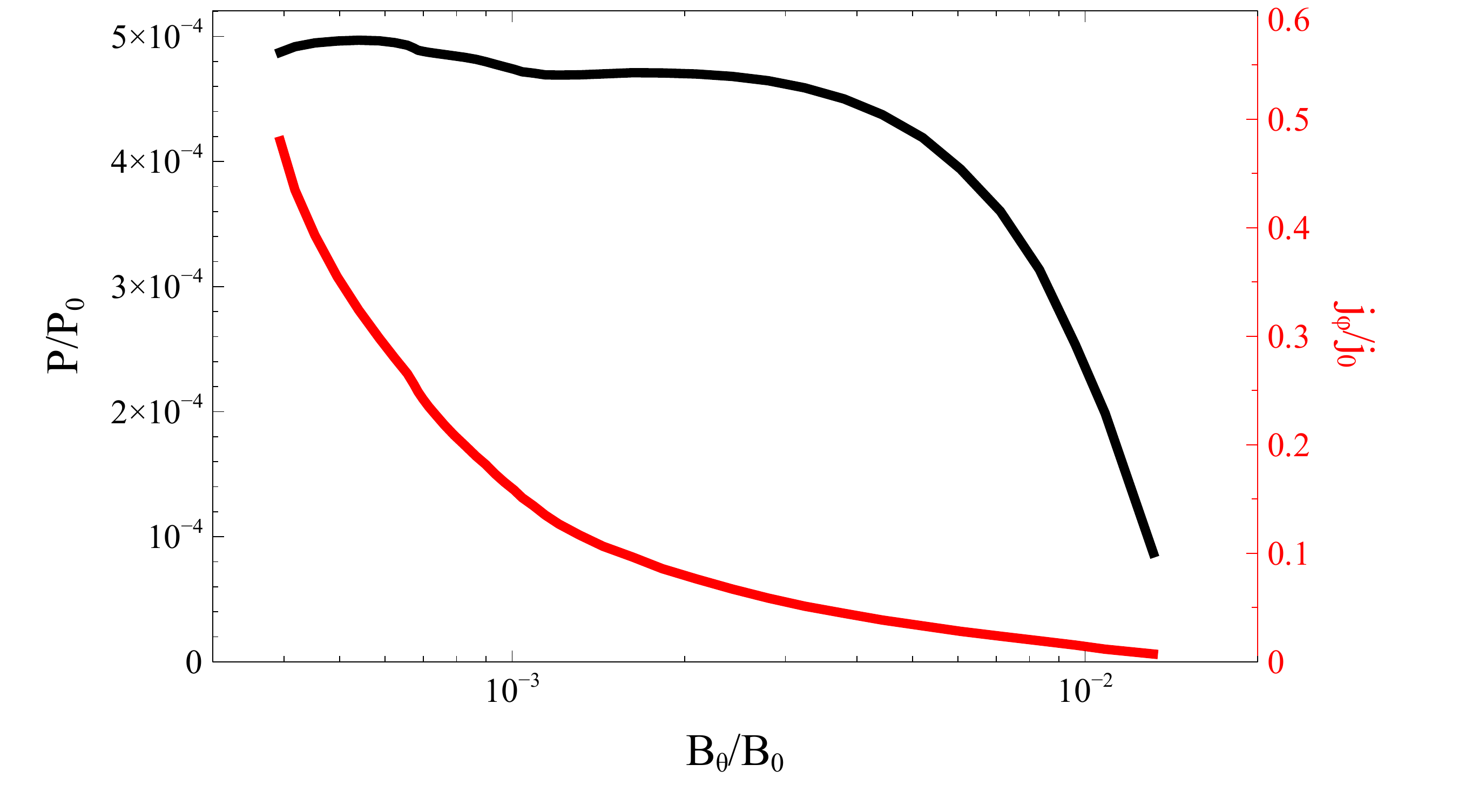}
\caption{\label{streamer:04} Profiles of the current density and plasma pressure from Fig. \ref{streamer:03} plotted as a function of $B_\theta$. See details of the normalizations and units in the caption of Fig. \ref{streamer:02}.}
\end{figure}

Similar to Fig. \ref{magnetotail:01}(d,e), Figure \ref{streamer:04} shows $j_\varphi$, $P$ as a function of $B_\theta$. Figures \ref{streamer:04}, \ref{magnetotail:02} show an important similarity of the thin current sheet formation in the helmet streamer system and in the Earth’s magnetotail. In both systems the strong increase of the current density is not associated with the strong plasma pressure increase (equatorial $P\approx B_L^2/8\pi$ in the magnetotail), i.e., the current sheet thinning is characterized by an internal reconfiguration of the cross-sheet $\theta$-profiles ($z$-profiles) of plasma pressure without significant {\it loading} of magnetic fluxes in current sheet boundaries. Therefore, the pre-reconnection thin current sheet configuration in the helmet streamer, and the corresponding reconnection properties of this current sheet, are controlled by the internal nonlinear plasma dynamics rather than by external (boundary) variations. The same conclusion for the magnetotail has been acknowledged in series of models, which show the thin current sheet develops as a result of internal reconfiguration of the plasma and magnetic field, without strong variations of the magnetic field pressure in current sheet boundaries \citep[see][]{Birn96, Birn04MHD, Hsieh&Otto15}.

\section{Conclusions}\label{sec:conclusions}
In this study we compare details of the thin current sheet formation in Earth's magnetotail (as derived from in-situ spacecraft observations) and in the helmet streamer magnetic field configuration (as derived from the MHD simulation). This comparison reveals interesting similarities of the current sheet thinning in both systems:
\begin{itemize}
\item Despite the different relation between thermal plasma pressure $P$ and dynamic pressure $\rho v^2/2$ for the magnetotail ($P\gg \rho v^2/2$) and helmet streamer configuration ($P\ll \rho v^2/2$), in both systems the current sheet thinning is controlled by the fine balance of the magnetic field line tension force $\sim [({\bf B}\nabla){\bf B}]$ and plasma pressure gradient $\sim \nabla P$ around the narrow, near-equatorial layer of strong current density. Before the reconnection onset the dynamic plasma pressure contribution to the current sheet stress balance is not important. 
\item The current sheet thinning is associated with the current density increase and equatorial magnetic field decrease (keeping $j_\varphi B_\theta \approx const$), such that the magnetic field line tension force does not change much. Thus, there is no significant increase of the plasma pressure magnitude, and the thin current sheet formation is characterized by the internal reconfiguration of the current density and plasma pressure along $z$.  
\end{itemize}
Similarities of these two systems suggest that many kinetic effects (Hall fields, collisionless conductivity, electron anisotropy, and ion nongyrotropy), typical for the thin magnetotail current sheet \citep[see discussion in reviews by][]{Artemyev&Zelenyi13,Sitnov19,Runov21:jastp}, may be valid for the thin current sheet formation in the helmet streamer configuration. Inclusion of such effects for large-scale systems as helmet streamers cannot be performed in full particle-in-cell simulations. However, some of these effects can be well reproduced in global hybrid simulations \citep[see][]{ Lin17:kaw_modelling,Lu16:cs, Lu17:ti, Lu19:jgr:cs}, which cover spatial scales of planetary magnetospheres \citep[e.g.,][]{Lin14:hybrid_code,Runov21:angeo},  and may be applied to spatial scales of the helmet streamer systems. 

\section*{Acknowledgements}
This work is supported NASA DRIVE SCIENCE CENTER HERMES grant No. 80NSSC20K0604 (A.A., V.R., A.R., Y.N., M.V.) and by ERC SLOW SOURCE DLV-819189 (V.R.). I.Z. acknowledges the support from the Russian Science Foundation grant No. 17-72-20134, Y.N. was supported by NASA grant 80NSSC18K0657 and 80NSSC21K1321, NSF grant AGS-1907698 and AGS-2100975, and the "Magnetotail Dipolarizations: Archimedes Force or Ideal Collapse?" ISSI team. 

We acknowledge NASA contract NAS5-02099 for use of THEMIS data. We would like to thank C. W. Carlson and J. P. McFadden for use of ESA data, D.E. Larson and R.P. Lin for use of SST data, K. H. Glassmeier, U. Auster, and W. Baumjohann for the use of FGM data provided under the lead of the Technical University of Braunschweig and with a financial support through the German Ministry for Economy and Technology and the German Aerospace Center (DLR) under contract 50 OC 0302. THEMIS data were downloaded from http://themis.ssl.berkeley.edu/. Data access and processing was done using SPEDAS V4.1, see \citet{Angelopoulos19}.


\begin{thebibliography}{}
\expandafter\ifx\csname natexlab\endcsname\relax\def\natexlab#1{#1}\fi
\providecommand{\url}[1]{\href{#1}{#1}}
\providecommand{\dodoi}[1]{doi:~\href{http://doi.org/#1}{\nolinkurl{#1}}}
\providecommand{\doeprint}[1]{\href{http://ascl.net/#1}{\nolinkurl{http://ascl.net/#1}}}
\providecommand{\doarXiv}[1]{\href{https://arxiv.org/abs/#1}{\nolinkurl{https://arxiv.org/abs/#1}}}

\bibitem[{{Airapetian} {et~al.}(2011){Airapetian}, {Ofman}, {Sittler}, \&
  {Kramar}}]{Airapetian11}
{Airapetian}, V., {Ofman}, L., {Sittler}, E.~C., \& {Kramar}, M. 2011, \apj,
  728, 67, \dodoi{10.1088/0004-637X/728/1/67}

\bibitem[{{Angelopoulos}(2008)}]{Angelopoulos08:ssr}
{Angelopoulos}, V. 2008, \ssr, 141, 5, \dodoi{10.1007/s11214-008-9336-1}

\bibitem[{{Angelopoulos} {et~al.}(2008{\natexlab{a}}){Angelopoulos},
  {McFadden}, {Larson}, {Carlson}, {Mende}, {Frey}, {Phan}, {Sibeck},
  {Glassmeier}, {Auster}, {Donovan}, {Mann}, {Rae}, {Russell}, {Runov}, {Zhou},
  \& {Kepko}}]{Angelopoulos08}
{Angelopoulos}, V., {McFadden}, J.~P., {Larson}, D., {et~al.}
  2008{\natexlab{a}}, Science, 321, 931, \dodoi{10.1126/science.1160495}

\bibitem[{{Angelopoulos} {et~al.}(2008{\natexlab{b}}){Angelopoulos}, {Sibeck},
  {Carlson}, {McFadden}, {Larson}, {Lin}, {Bonnell}, {Mozer}, {Ergun}, {Cully},
  {Glassmeier}, {Auster}, {Roux}, {Lecontel}, {Frey}, {Phan}, {Mende}, {Frey},
  {Donovan}, {Russell}, {Strangeway}, {Liu}, {Mann}, {Rae}, {Raeder}, {Li},
  {Liu}, {Singer}, {Sergeev}, {Apatenkov}, {Parks}, {Fillingim}, \&
  {Sigwarth}}]{Angelopoulos08:sst}
{Angelopoulos}, V., {Sibeck}, D., {Carlson}, C.~W., {et~al.}
  2008{\natexlab{b}}, \ssr, 141, 453, \dodoi{10.1007/s11214-008-9378-4}

\bibitem[{{Angelopoulos} {et~al.}(2019){Angelopoulos}, {Cruce}, {Drozdov},
  {Grimes}, {Hatzigeorgiu}, {King}, {Larson}, {Lewis}, {McTiernan}, {Roberts},
  {Russell}, {Hori}, {Kasahara}, {Kumamoto}, {Matsuoka}, {Miyashita},
  {Miyoshi}, {Shinohara}, {Teramoto}, {Faden}, {Halford}, {McCarthy}, {Millan},
  {Sample}, {Smith}, {Woodger}, {Masson}, {Narock}, {Asamura}, {Chang},
  {Chiang}, {Kazama}, {Keika}, {Matsuda}, {Segawa}, {Seki}, {Shoji}, {Tam},
  {Umemura}, {Wang}, {Wang}, {Redmon}, {Rodriguez}, {Singer}, {Vandegriff},
  {Abe}, {Nose}, {Shinbori}, {Tanaka}, {UeNo}, {Andersson}, {Dunn}, {Fowler},
  {Halekas}, {Hara}, {Harada}, {Lee}, {Lillis}, {Mitchell}, {Argall},
  {Bromund}, {Burch}, {Cohen}, {Galloy}, {Giles}, {Jaynes}, {Le Contel}, {Oka},
  {Phan}, {Walsh}, {Westlake}, {Wilder}, {Bale}, {Livi}, {Pulupa},
  {Whittlesey}, {DeWolfe}, {Harter}, {Lucas}, {Auster}, {Bonnell}, {Cully},
  {Donovan}, {Ergun}, {Frey}, {Jackel}, {Keiling}, {Korth}, {McFadden},
  {Nishimura}, {Plaschke}, {Robert}, {Turner}, {Weygand}, {Candey}, {Johnson},
  {Kovalick}, {Liu}, {McGuire}, {Breneman}, {Kersten}, \&
  {Schroeder}}]{Angelopoulos19}
{Angelopoulos}, V., {Cruce}, P., {Drozdov}, A., {et~al.} 2019, \ssr, 215, 9,
  \dodoi{10.1007/s11214-018-0576-4}

\bibitem[{{Artemyev} {et~al.}(2016){Artemyev}, {Angelopoulos}, {Runov}, \&
  {Petrukovich}}]{Artemyev16:jgr:thinning}
{Artemyev}, A.~V., {Angelopoulos}, V., {Runov}, A., \& {Petrukovich}, A.~A.
  2016, \jgr, 121, 6718, \dodoi{10.1002/2016JA022779}

\bibitem[{{Artemyev} {et~al.}(2019){Artemyev}, {Angelopoulos}, {Runov}, \&
  {Petrukovich}}]{Artemyev19:jgr:globalview}
---. 2019, Journal of Geophysical Research (Space Physics), 124, 264,
  \dodoi{10.1029/2018JA026113}

\bibitem[{{Artemyev} {et~al.}(2010){Artemyev}, {Petrukovich}, {Nakamura}, \&
  {Zelenyi}}]{Artemyev10:jgr}
{Artemyev}, A.~V., {Petrukovich}, A.~A., {Nakamura}, R., \& {Zelenyi}, L.~M.
  2010, J. Geophys. Res., 115, A12255, \dodoi{10.1029/2010JA015702}

\bibitem[{{Artemyev} {et~al.}(2011){Artemyev}, {Petrukovich}, {Nakamura}, \&
  {Zelenyi}}]{Artemyev11:jgr}
---. 2011, J. Geophys. Res., 116, A0923, \dodoi{10.1029/2011JA016801}

\bibitem[{{Artemyev} \& {Zelenyi}(2013)}]{Artemyev&Zelenyi13}
{Artemyev}, A.~V., \& {Zelenyi}, L.~M. 2013, Space Sci. Rev., 178, 419,
  \dodoi{10.1007/s11214-012-9954-5}

\bibitem[{{Artemyev} {et~al.}(2021){Artemyev}, {Lu}, {El-Alaoui}, {Lin},
  {Angelopoulos}, {Zhang}, {Runov}, {Vasko}, {Zelenyi}, \&
  {Russell}}]{Artemyev21:grl}
{Artemyev}, A.~V., {Lu}, S., {El-Alaoui}, M., {et~al.} 2021, \grl, 48, e92153,
  \dodoi{10.1029/2020GL092153}

\bibitem[{{Aschwanden}(2002)}]{Aschwanden02}
{Aschwanden}, M.~J. 2002, \ssr, 101, 1, \dodoi{10.1023/A:1019712124366}

\bibitem[{{Auster} {et~al.}(2008){Auster}, {Glassmeier}, {Magnes}, {Aydogar},
  {Baumjohann}, {Constantinescu}, {Fischer}, {Fornacon}, {Georgescu}, {Harvey},
  {Hillenmaier}, {Kroth}, {Ludlam}, {Narita}, {Nakamura}, {Okrafka},
  {Plaschke}, {Richter}, {Schwarzl}, {Stoll}, {Valavanoglou}, \&
  {Wiedemann}}]{Auster08:THEMIS}
{Auster}, H.~U., {Glassmeier}, K.~H., {Magnes}, W., {et~al.} 2008, \ssr, 141,
  235, \dodoi{10.1007/s11214-008-9365-9}

\bibitem[{{Baker} {et~al.}(1996){Baker}, {Pulkkinen}, {Angelopoulos},
  {Baumjohann}, \& {McPherron}}]{Baker96}
{Baker}, D.~N., {Pulkkinen}, T.~I., {Angelopoulos}, V., {Baumjohann}, W., \&
  {McPherron}, R.~L. 1996, \jgr, 101, 12975, \dodoi{10.1029/95JA03753}

\bibitem[{{Baumjohann} {et~al.}(2007){Baumjohann}, {Roux}, {Le Contel},
  {Nakamura}, {Birn}, {Hoshino}, {Lui}, {Owen}, {Sauvaud}, {Vaivads},
  {Fontaine}, \& {Runov}}]{Baumjohann07}
{Baumjohann}, W., {Roux}, A., {Le Contel}, O., {et~al.} 2007, Annales
  Geophysicae, 25, 1365

\bibitem[{{Bavassano} {et~al.}(1997){Bavassano}, {Woo}, \&
  {Bruno}}]{Bavassano97}
{Bavassano}, B., {Woo}, R., \& {Bruno}, R. 1997, \grl, 24, 1655,
  \dodoi{10.1029/97GL01630}

\bibitem[{{Birn}(1991)}]{Birn91:pop}
{Birn}, J. 1991, Physics of Fluids B, 3, 479, \dodoi{10.1063/1.859891}

\bibitem[{{Birn}(1992)}]{Birn92}
---. 1992, \jgr, 97, 16817, \dodoi{10.1029/92JA01527}

\bibitem[{{Birn} {et~al.}(2004){Birn}, {Dorelli}, {Hesse}, \&
  {Schindler}}]{Birn04MHD}
{Birn}, J., {Dorelli}, J.~C., {Hesse}, M., \& {Schindler}, K. 2004, J. Geophys.
  Res., 109, 2215, \dodoi{10.1029/2003JA010275}

\bibitem[{Birn \& Hesse(2005)}]{Birn&Hesse05}
Birn, J., \& Hesse, M. 2005, Annales Geophysicae, 23, 3365,
  \dodoi{10.5194/angeo-23-3365-2005}

\bibitem[{{Birn} \& {Hesse}(2009)}]{Birn&Hesse09}
{Birn}, J., \& {Hesse}, M. 2009, Annales Geophysicae, 27, 1067,
  \dodoi{10.5194/angeo-27-1067-2009}

\bibitem[{{Birn} \& {Hesse}(2014)}]{Birn&Hesse14}
---. 2014, \jgr, 119, 290, \dodoi{10.1002/2013JA019354}

\bibitem[{{Birn} {et~al.}(1996){Birn}, {Hesse}, \& {Schindler}}]{Birn96}
{Birn}, J., {Hesse}, M., \& {Schindler}, K. 1996, \jgr, 101, 12939,
  \dodoi{10.1029/96JA00611}

\bibitem[{{Birn} {et~al.}(1998){Birn}, {Hesse}, \& {Schindler}}]{Birn98:cs}
---. 1998, \jgr, 103, 6843, \dodoi{10.1029/97JA03602}

\bibitem[{{Birn} \& {Priest}(2007)}]{bookBirn&Priest07}
{Birn}, J., \& {Priest}, E.~R. 2007, {Reconnection of magnetic fields :
  magnetohydrodynamics and collisionless theory and observations}, ed. {Birn,
  J.~\& Priest, E.~R.}

\bibitem[{{Birn} {et~al.}(1977){Birn}, {Sommer}, \& {Schindler}}]{Birn77}
{Birn}, J., {Sommer}, R.~R., \& {Schindler}, K. 1977, J. Geophys. Res., 82,
  147, \dodoi{10.1029/JA082i001p00147}

\bibitem[{{Biskamp}(2000)}]{bookBiskamp00}
{Biskamp}, D. 2000, {Magnetic Reconnection in Plasmas}

\bibitem[{{Cheng}(1992)}]{Cheng92}
{Cheng}, C.~Z. 1992, \jgr, 97, 1497, \dodoi{10.1029/91JA02433}

\bibitem[{{Cuperman} {et~al.}(1993){Cuperman}, {Bruma}, {Detman}, \&
  {Dryer}}]{Cuperman93}
{Cuperman}, S., {Bruma}, C., {Detman}, T., \& {Dryer}, M. 1993, \apj, 404, 356,
  \dodoi{10.1086/172285}

\bibitem[{{Cuperman} {et~al.}(1995){Cuperman}, {Bruma}, {Dryer}, \&
  {Semel}}]{Cuperman95}
{Cuperman}, S., {Bruma}, C., {Dryer}, M., \& {Semel}, M. 1995, \aap, 299, 389

\bibitem[{{Cuperman} {et~al.}(1992){Cuperman}, {Detman}, {Bruma}, \&
  {Dryer}}]{Cuperman92}
{Cuperman}, S., {Detman}, T.~R., {Bruma}, C., \& {Dryer}, M. 1992, \aap, 265,
  785

\bibitem[{{Cuperman} {et~al.}(1990){Cuperman}, {Ofman}, \&
  {Dryer}}]{Cuperman90}
{Cuperman}, S., {Ofman}, L., \& {Dryer}, M. 1990, \apj, 350, 846,
  \dodoi{10.1086/168436}

\bibitem[{{Dahlburg} \& {Karpen}(1995)}]{Dahlburg&Karpen95}
{Dahlburg}, R.~B., \& {Karpen}, J.~T. 1995, \jgr, 100, 23489,
  \dodoi{10.1029/95JA02496}

\bibitem[{{Dunlop} {et~al.}(2002){Dunlop}, {Balogh}, {Glassmeier}, \&
  {Robert}}]{Dunlop02}
{Dunlop}, M.~W., {Balogh}, A., {Glassmeier}, K.-H., \& {Robert}, P. 2002, \jgr,
  107, 1384, \dodoi{10.1029/2001JA005088}

\bibitem[{{Einaudi}(1999)}]{Einaudi99}
{Einaudi}, G. 1999, Plasma Physics and Controlled Fusion, 41, A293,
  \dodoi{10.1088/0741-3335/41/3A/023}

\bibitem[{{Endeve} {et~al.}(2004){Endeve}, {Holzer}, \& {Leer}}]{Endeve04}
{Endeve}, E., {Holzer}, T.~E., \& {Leer}, E. 2004, \apj, 603, 307,
  \dodoi{10.1086/381239}

\bibitem[{{Escoubet} {et~al.}(2001){Escoubet}, {Fehringer}, \&
  {Goldstein}}]{Escoubet01}
{Escoubet}, C.~P., {Fehringer}, M., \& {Goldstein}, M. 2001, Annales
  Geophysicae, 19, 1197, \dodoi{10.5194/angeo-19-1197-2001}

\bibitem[{{Eselevich} \& {Filippov}(1988)}]{Eselevich&Filippov88}
{Eselevich}, V.~G., \& {Filippov}, M.~A. 1988, \planss, 36, 105,
  \dodoi{10.1016/0032-0633(88)90046-3}

\bibitem[{{Feng} {et~al.}(2013){Feng}, {Inhester}, \& {Gan}}]{Feng13}
{Feng}, L., {Inhester}, B., \& {Gan}, W.~Q. 2013, \apj, 774, 141,
  \dodoi{10.1088/0004-637X/774/2/141}

\bibitem[{{Gonzalez} \& {Parker}(2016)}]{book:Gonzalez&Parker}
{Gonzalez}, W., \& {Parker}, E. 2016, {Magnetic Reconnection}, Vol. 427,
  \dodoi{10.1007/978-3-319-26432-5}

\bibitem[{{Gopalswamy}(2003)}]{Gopalswamy03}
{Gopalswamy}, N. 2003, Advances in Space Research, 31, 869,
  \dodoi{10.1016/S0273-1177(02)00888-8}

\bibitem[{{Gosling} {et~al.}(1981){Gosling}, {Borrini}, {Asbridge}, {Bame},
  {Feldman}, \& {Hansen}}]{Gosling81}
{Gosling}, J.~T., {Borrini}, G., {Asbridge}, J.~R., {et~al.} 1981, \jgr, 86,
  5438, \dodoi{10.1029/JA086iA07p05438}

\bibitem[{{Guo} \& {Wu}(1998)}]{Guo&Wu98}
{Guo}, W.~P., \& {Wu}, S.~T. 1998, \apj, 494, 419, \dodoi{10.1086/305196}

\bibitem[{{Guo} {et~al.}(1996){Guo}, {Wu}, \& {Tandberg-Hanssen}}]{Guo96}
{Guo}, W.~P., {Wu}, S.~T., \& {Tandberg-Hanssen}, E. 1996, \apj, 469, 944,
  \dodoi{10.1086/177841}

\bibitem[{{Hodgson} \& {Neukirch}(2015)}]{Hodgson&Neukirch15}
{Hodgson}, J.~D.~B., \& {Neukirch}, T. 2015, Geophysical and Astrophysical
  Fluid Dynamics, 109, 524, \dodoi{10.1080/03091929.2015.1081188}

\bibitem[{{Hsieh} \& {Otto}(2015)}]{Hsieh&Otto15}
{Hsieh}, M.-S., \& {Otto}, A. 2015, \jgr, 120, 4264,
  \dodoi{10.1002/2014JA020925}

\bibitem[{{Jackman} {et~al.}(2014){Jackman}, {Arridge}, {Andr{\'e}}, {Bagenal},
  {Birn}, {Freeman}, {Jia}, {Kidder}, {Milan}, {Radioti}, {Slavin}, {Vogt},
  {Volwerk}, \& {Walsh}}]{Jackman14}
{Jackman}, C.~M., {Arridge}, C.~S., {Andr{\'e}}, N., {et~al.} 2014, \ssr, 182,
  85, \dodoi{10.1007/s11214-014-0060-8}

\bibitem[{{Kivelson} {et~al.}(2005){Kivelson}, {McPherron}, {Thompson},
  {Khurana}, {Weygand}, \& {Balogh}}]{Kivelson05}
{Kivelson}, M.~G., {McPherron}, R.~L., {Thompson}, S., {et~al.} 2005, Advances
  in Space Research, 36, 1818, \dodoi{10.1016/j.asr.2004.03.024}

\bibitem[{{Korreck} {et~al.}(2020){Korreck}, {Szabo}, {Nieves Chinchilla},
  {Lavraud}, {Luhmann}, {Niembro}, {Higginson}, {Alzate}, {Wallace}, {Paulson},
  {Rouillard}, {Kouloumvakos}, {Poirier}, {Kasper}, {Case}, {Stevens}, {Bale},
  {Pulupa}, {Whittlesey}, {Livi}, {Goetz}, {Larson}, {Malaspina}, {Morgan},
  {Narock}, {Schwadron}, {Bonnell}, {Harvey}, \& {Wygant}}]{Korreck20}
{Korreck}, K.~E., {Szabo}, A., {Nieves Chinchilla}, T., {et~al.} 2020, \apjs,
  246, 69, \dodoi{10.3847/1538-4365/ab6ff9}

\bibitem[{{Lapenta} \& {Knoll}(2003)}]{Lapenta&Knoll03}
{Lapenta}, G., \& {Knoll}, D.~A. 2003, \solphys, 214, 107,
  \dodoi{10.1023/A:1024036917505}

\bibitem[{{Lee} {et~al.}(2021){Lee}, {Cho}, {An}, {Lee}, {Seough}, {Kim}, \&
  {Kumar}}]{Lee21:streamers}
{Lee}, J.-O., {Cho}, K.-S., {An}, J., {et~al.} 2021, \apjl, 920, L6,
  \dodoi{10.3847/2041-8213/ac2422}

\bibitem[{{Liewer} {et~al.}(2021){Liewer}, {Qiu}, {Vourlidas}, {Hall}, \&
  {Penteado}}]{Liewer21}
{Liewer}, P.~C., {Qiu}, J., {Vourlidas}, A., {Hall}, J.~R., \& {Penteado}, P.
  2021, \aap, 650, A32, \dodoi{10.1051/0004-6361/202039641}

\bibitem[{{Lin} {et~al.}(2014){Lin}, {Wang}, {Lu}, {Perez}, \&
  {Lu}}]{Lin14:hybrid_code}
{Lin}, Y., {Wang}, X.~Y., {Lu}, S., {Perez}, J.~D., \& {Lu}, Q. 2014, \jgr,
  119, 7413, \dodoi{10.1002/2014JA020005}

\bibitem[{{Lin} {et~al.}(2017){Lin}, {Wing}, {Johnson}, {Wang}, {Perez}, \&
  {Cheng}}]{Lin17:kaw_modelling}
{Lin}, Y., {Wing}, S., {Johnson}, J.~R., {et~al.} 2017, \grl, 44, 5892,
  \dodoi{10.1002/2017GL073957}

\bibitem[{{Linker} {et~al.}(2001){Linker}, {Lionello}, {Miki{\'c}}, \&
  {Amari}}]{Linker01}
{Linker}, J.~A., {Lionello}, R., {Miki{\'c}}, Z., \& {Amari}, T. 2001, \jgr,
  106, 25165, \dodoi{10.1029/2000JA004020}

\bibitem[{{Liu} {et~al.}(2014){Liu}, {Birn}, {Daughton}, {Hesse}, \&
  {Schindler}}]{Liu14:CS}
{Liu}, Y.-H., {Birn}, J., {Daughton}, W., {Hesse}, M., \& {Schindler}, K. 2014,
  \jgr, 119, 9773, \dodoi{10.1002/2014JA020492}

\bibitem[{{Lu} {et~al.}(2017){Lu}, {Artemyev}, {Angelopoulos}, {Lin}, \&
  {Wang}}]{Lu17:ti}
{Lu}, S., {Artemyev}, A.~V., {Angelopoulos}, V., {Lin}, Y., \& {Wang}, X.~Y.
  2017, \jgr, 122, 8295, \dodoi{10.1002/2017JA024209}

\bibitem[{{Lu} {et~al.}(2016){Lu}, {Lin}, {Angelopoulos}, {Artemyev},
  {Pritchett}, {Lu}, \& {Wang}}]{Lu16:cs}
{Lu}, S., {Lin}, Y., {Angelopoulos}, V., {et~al.} 2016, \jgr, 121, 11,
  \dodoi{10.1002/2016JA023325}

\bibitem[{{Lu} {et~al.}(2019){Lu}, {Artemyev}, {Angelopoulos}, {Lin}, {Zhang},
  {Liu}, {Avanov}, {Giles}, {Russell}, \& {Strangeway}}]{Lu19:jgr:cs}
{Lu}, S., {Artemyev}, A.~V., {Angelopoulos}, V., {et~al.} 2019, Journal of
  Geophysical Research (Space Physics), 124, 1052, \dodoi{10.1029/2018JA026202}

\bibitem[{{McFadden} {et~al.}(2008){McFadden}, {Carlson}, {Larson}, {Ludlam},
  {Abiad}, {Elliott}, {Turin}, {Marckwordt}, \&
  {Angelopoulos}}]{McFadden08:THEMIS}
{McFadden}, J.~P., {Carlson}, C.~W., {Larson}, D., {et~al.} 2008, \ssr, 141,
  277, \dodoi{10.1007/s11214-008-9440-2}

\bibitem[{{Mignone} {et~al.}(2007){Mignone}, {Bodo}, {Massaglia}, {Matsakos},
  {Tesileanu}, {Zanni}, \& {Ferrari}}]{Mignone07}
{Mignone}, A., {Bodo}, G., {Massaglia}, S., {et~al.} 2007, \apjs, 170, 228,
  \dodoi{10.1086/513316}

\bibitem[{{Nakamura} {et~al.}(2002){Nakamura}, {Baumjohann}, {Klecker},
  {Bogdanova}, {Balogh}, {R{\`e}me}, {Bosqued}, {Dandouras}, {Sauvaud},
  {Glassmeier}, {Kistler}, {Mouikis}, {Zhang}, {Eichelberger}, \&
  {Runov}}]{Nakamura02}
{Nakamura}, R., {Baumjohann}, W., {Klecker}, B., {et~al.} 2002, \grl, 29,
  200000, \dodoi{10.1029/2002GL015763}

\bibitem[{{Nakamura} {et~al.}(2009){Nakamura}, {Retin{\`o}}, {Baumjohann},
  {Volwerk}, {Erkaev}, {Klecker}, {Lucek}, {Dandouras}, {Andr{\'e}}, \&
  {Khotyaintsev}}]{Nakamura09}
{Nakamura}, R., {Retin{\`o}}, A., {Baumjohann}, W., {et~al.} 2009, Annales
  Geophysicae, 27, 1743

\bibitem[{{Neukirch}(1995{\natexlab{a}})}]{Neukirch95}
{Neukirch}, T. 1995{\natexlab{a}}, \aap, 301, 628

\bibitem[{{Neukirch}(1995{\natexlab{b}})}]{Neukirch95:pop}
---. 1995{\natexlab{b}}, Physics of Plasmas, 2, 4389, \dodoi{10.1063/1.870995}

\bibitem[{{Neukirch}(1997)}]{Neukirch97}
---. 1997, \aap, 325, 847

\bibitem[{{Nickeler} \& {Wiegelmann}(2010)}]{Nickeler&Wiegelmann10}
{Nickeler}, D.~H., \& {Wiegelmann}, T. 2010, Annales Geophysicae, 28, 1523,
  \dodoi{10.5194/angeo-28-1523-2010}

\bibitem[{{Nieves-Chinchilla} {et~al.}(2020){Nieves-Chinchilla}, {Szabo},
  {Korreck}, {Alzate}, {Balmaceda}, {Lavraud}, {Paulson}, {Narock}, {Wallace},
  {Jian}, {Luhmann}, {Morgan}, {Higginson}, {Arge}, {Bale}, {Case}, {Dudfok de
  Wit}, {Giacalone}, {Goetz}, {Harvey}, {Jones-Melosky}, {Kasper}, {Larson},
  {Livi}, {McComas}, {MacDowall}, {Malaspina}, {Pulupa}, {Raouafi},
  {Schwadron}, {Stevens}, \& {Whittlesey}}]{NievesChinchilla20}
{Nieves-Chinchilla}, T., {Szabo}, A., {Korreck}, K.~E., {et~al.} 2020, \apjs,
  246, 63, \dodoi{10.3847/1538-4365/ab61f5}

\bibitem[{{Nishimura} \& {Lyons}(2016)}]{Nishimura&Lyons16:flows}
{Nishimura}, Y., \& {Lyons}, L.~R. 2016, Journal of Geophysical Research (Space
  Physics), 121, 1327, \dodoi{10.1002/2015JA022128}

\bibitem[{{Ofman} {et~al.}(2011){Ofman}, {Abbo}, \& {Giordano}}]{Ofman11}
{Ofman}, L., {Abbo}, L., \& {Giordano}, S. 2011, \apj, 734, 30,
  \dodoi{10.1088/0004-637X/734/1/30}

\bibitem[{{Ofman} {et~al.}(2015){Ofman}, {Provornikova}, {Abbo}, \&
  {Giordano}}]{Ofman15}
{Ofman}, L., {Provornikova}, E., {Abbo}, L., \& {Giordano}, S. 2015, Annales
  Geophysicae, 33, 47, \dodoi{10.5194/angeo-33-47-2015}

\bibitem[{{Parker}(1994)}]{bookParker94}
{Parker}, E.~N. 1994, Spontaneous current sheets in magnetic fields: with
  applications to stellar x-rays.~ International Series in Astronomy and
  Astrophysics, Vol.~1.~ New York : Oxford University Press, 1994., 1

\bibitem[{{Petrukovich} {et~al.}(2013){Petrukovich}, {Artemyev}, {Nakamura},
  {Panov}, \& {Baumjohann}}]{Petrukovich13}
{Petrukovich}, A.~A., {Artemyev}, A.~V., {Nakamura}, R., {Panov}, E.~V., \&
  {Baumjohann}, W. 2013, \jgr, 118, 5720, \dodoi{10.1002/jgra.50550}

\bibitem[{{Petrukovich} {et~al.}(2015){Petrukovich}, {Artemyev}, {Vasko},
  {Nakamura}, \& {Zelenyi}}]{Petrukovich15:ssr}
{Petrukovich}, A.~A., {Artemyev}, A.~V., {Vasko}, I.~Y., {Nakamura}, R., \&
  {Zelenyi}, L.~M. 2015, \ssr, 188, 311, \dodoi{10.1007/s11214-014-0126-7}

\bibitem[{{Petrukovich} {et~al.}(2009){Petrukovich}, {Baumjohann}, {Nakamura},
  \& {R{\`e}me}}]{Petrukovich09}
{Petrukovich}, A.~A., {Baumjohann}, W., {Nakamura}, R., \& {R{\`e}me}, H. 2009,
  J. Geophys. Res., 114, 9203, \dodoi{10.1029/2009JA014064}

\bibitem[{{Petrukovich} {et~al.}(2007){Petrukovich}, {Baumjohann}, {Nakamura},
  {Runov}, {Balogh}, \& {R{\`e}me}}]{Petrukovich07}
{Petrukovich}, A.~A., {Baumjohann}, W., {Nakamura}, R., {et~al.} 2007, \jgr,
  112, 10213, \dodoi{10.1029/2007JA012349}

\bibitem[{{Petrukovich} {et~al.}(1999){Petrukovich}, {Mukai}, {Kokubun},
  {Romanov}, {Saito}, {Yamamoto}, \& {Zelenyi}}]{Petrukovich99}
{Petrukovich}, A.~A., {Mukai}, T., {Kokubun}, S., {et~al.} 1999, \jgr, 104,
  4501, \dodoi{10.1029/98JA02418}

\bibitem[{{Pneuman}(1972)}]{Pneuman72}
{Pneuman}, G.~W. 1972, \solphys, 23, 223, \dodoi{10.1007/BF00153906}

\bibitem[{{Pneuman} \& {Kopp}(1971)}]{Pneuman&Kopp71}
{Pneuman}, G.~W., \& {Kopp}, R.~A. 1971, \solphys, 18, 258,
  \dodoi{10.1007/BF00145940}

\bibitem[{{Priest}(2016)}]{bookPriest16}
{Priest}, E. 2016, in Astrophysics and Space Science Library, Vol. 427,
  Astrophysics and Space Science Library, ed. W.~{Gonzalez} \& E.~{Parker},
  101, \dodoi{10.1007/978-3-319-26432-5-3}

\bibitem[{{Priest}(1985)}]{Priest85}
{Priest}, E.~R. 1985, Reports on Progress in Physics, 48, 955,
  \dodoi{10.1088/0034-4885/48/7/002}

\bibitem[{{Priest} \& {Forbes}(2002)}]{Priest&Forbes02}
{Priest}, E.~R., \& {Forbes}, T.~G. 2002, \aapr, 10, 313,
  \dodoi{10.1007/s001590100013}

\bibitem[{{Pritchett} \& {Coroniti}(1994)}]{Pritchett&Coroniti94}
{Pritchett}, P.~L., \& {Coroniti}, F.~V. 1994, \grl, 21, 1587,
  \dodoi{10.1029/94GL01364}

\bibitem[{{Pritchett} \& {Coroniti}(1995)}]{Pritchett&Coroniti95}
---. 1995, \jgr, 100, 23551, \dodoi{10.1029/95JA02540}

\bibitem[{{Pritchett} \& {Lu}(2018)}]{Pritchett&Lu18}
{Pritchett}, P.~L., \& {Lu}, S. 2018, \jgr, 123, 2787,
  \dodoi{10.1002/2017JA025094}

\bibitem[{{Rast{\"a}tter} {et~al.}(1999){Rast{\"a}tter}, {Hesse}, \&
  {Schindler}}]{Rastaetter99}
{Rast{\"a}tter}, L., {Hesse}, M., \& {Schindler}, K. 1999, \jgr, 104, 12301,
  \dodoi{10.1029/1999JA900138}

\bibitem[{{Reeves} {et~al.}(2008){Reeves}, {Guild}, {Hughes}, {Korreck}, {Lin},
  {Raymond}, {Savage}, {Schwadron}, {Spence}, {Webb}, \&
  {Wiltberger}}]{Reeves08:solar}
{Reeves}, K.~K., {Guild}, T.~B., {Hughes}, W.~J., {et~al.} 2008, \jgr, 113, 0,
  \dodoi{10.1029/2008JA013049}

\bibitem[{{R{\'e}ville} {et~al.}(2020{\natexlab{a}}){R{\'e}ville}, {Velli},
  {Rouillard}, {Lavraud}, {Tenerani}, {Shi}, \& {Strugarek}}]{Reville20:ApJ}
{R{\'e}ville}, V., {Velli}, M., {Rouillard}, A.~P., {et~al.}
  2020{\natexlab{a}}, \apjl, 895, L20, \dodoi{10.3847/2041-8213/ab911d}

\bibitem[{{R{\'e}ville} {et~al.}(2020{\natexlab{b}}){R{\'e}ville}, {Velli},
  {Panasenco}, {Tenerani}, {Shi}, {Badman}, {Bale}, {Kasper}, {Stevens},
  {Korreck}, {Bonnell}, {Case}, {de Wit}, {Goetz}, {Harvey}, {Larson}, {Livi},
  {Malaspina}, {MacDowall}, {Pulupa}, \& {Whittlesey}}]{Reville20:ApJS}
{R{\'e}ville}, V., {Velli}, M., {Panasenco}, O., {et~al.} 2020{\natexlab{b}},
  \apjs, 246, 24, \dodoi{10.3847/1538-4365/ab4fef}

\bibitem[{{R{\'e}ville} {et~al.}(2021){R{\'e}ville}, {Fargette}, {Rouillard},
  {Lavraud}, {Velli}, {Strugarek}, {Parenti}, {Brun}, {Shi}, {Kouloumvakos},
  {Poirier}, {Pinto}, {Louarn}, {Fedorov}, {Owen}, {G{\'e}not}, {Horbury},
  {Laker}, {O'Brien}, {Angelini}, {Fauchon-Jones}, \& {Kasper}}]{Reville22:A&A}
{R{\'e}ville}, V., {Fargette}, N., {Rouillard}, A.~P., {et~al.} 2021, arXiv
  e-prints, arXiv:2112.07445.
\newblock \doarXiv{2112.07445}

\bibitem[{{Rich} {et~al.}(1972){Rich}, {Vasyliunas}, \& {Wolf}}]{Rich72}
{Rich}, F.~J., {Vasyliunas}, V.~M., \& {Wolf}, R.~A. 1972, J. Geophys. Res.,
  77, 4670, \dodoi{10.1029/JA077i025p04670}

\bibitem[{{Runov} {et~al.}(2021{\natexlab{a}}){Runov}, {Angelopoulos},
  {Artemyev}, {Weygand}, {Lu}, {Lin}, \& {Zhang}}]{Runov21:jastp}
{Runov}, A., {Angelopoulos}, V., {Artemyev}, A.~V., {et~al.}
  2021{\natexlab{a}}, Journal of Atmospheric and Solar-Terrestrial Physics,
  220, 105671, \dodoi{10.1016/j.jastp.2021.105671}

\bibitem[{{Runov} {et~al.}(2005){Runov}, {Sergeev}, {Nakamura}, {Baumjohann},
  {Zhang}, {Asano}, {Volwerk}, {V{\"o}r{\"o}s}, {Balogh}, \&
  {R{\`e}me}}]{Runov05:pss}
{Runov}, A., {Sergeev}, V.~A., {Nakamura}, R., {et~al.} 2005, \planss, 53, 237,
  \dodoi{10.1016/j.pss.2004.09.049}

\bibitem[{{Runov} {et~al.}(2006){Runov}, {Sergeev}, {Nakamura}, {Baumjohann},
  {Apatenkov}, {Asano}, {Takada}, {Volwerk}, {V{\"o}r{\"o}s}, {Zhang},
  {Sauvaud}, {R{\`e}me}, \& {Balogh}}]{Runov06}
---. 2006, Annales Geophysicae, 24, 247

\bibitem[{{Runov} {et~al.}(2009){Runov}, {Angelopoulos}, {Sitnov}, {Sergeev},
  {Bonnell}, {McFadden}, {Larson}, {Glassmeier}, \& {Auster}}]{Runov09grl}
{Runov}, A., {Angelopoulos}, V., {Sitnov}, M.~I., {et~al.} 2009, \grl, 36,
  L14106, \dodoi{10.1029/2009GL038980}

\bibitem[{{Runov} {et~al.}(2021{\natexlab{b}}){Runov}, {Grandin}, {Palmroth},
  {Battarbee}, {Ganse}, {Hietala}, {Hoilijoki}, {Kilpua}, {Pfau-Kempf},
  {Toledo-Redondo}, {Turc}, \& {Turner}}]{Runov21:angeo}
{Runov}, A., {Grandin}, M., {Palmroth}, M., {et~al.} 2021{\natexlab{b}},
  Annales Geophysicae, 39, 599, \dodoi{10.5194/angeo-39-599-2021}

\bibitem[{{Schindler} \& {Birn}(1978)}]{Schindler&Birn78}
{Schindler}, K., \& {Birn}, J. 1978, \physrep, 47, 109,
  \dodoi{10.1016/0370-1573(78)90016-9}

\bibitem[{{Schindler} \& {Birn}(1986)}]{Schindler&Birn86}
---. 1986, Space Science Reviews, 44, 307, \dodoi{10.1007/BF00200819}

\bibitem[{{Schindler} \& {Birn}(1999)}]{Schindler&Birn99}
---. 1999, \jgr, 104, 25001, \dodoi{10.1029/1999JA900258}

\bibitem[{{Schindler} {et~al.}(1983){Schindler}, {Birn}, \&
  {Janicke}}]{Schindler83}
{Schindler}, K., {Birn}, J., \& {Janicke}, L. 1983, \solphys, 87, 103,
  \dodoi{10.1007/BF00151164}

\bibitem[{{Shi} {et~al.}(2021){Shi}, {Artemyev}, {Velli}, \&
  {Tenerani}}]{Shi21:jgr:tearing}
{Shi}, C., {Artemyev}, A., {Velli}, M., \& {Tenerani}, A. 2021, Journal of
  Geophysical Research (Space Physics), 126, e29711,
  \dodoi{10.1029/2021JA029711}

\bibitem[{{Sitnov} {et~al.}(2009){Sitnov}, {Swisdak}, \& {Divin}}]{Sitnov09}
{Sitnov}, M.~I., {Swisdak}, M., \& {Divin}, A.~V. 2009, \jgr, 114, A04202,
  \dodoi{10.1029/2008JA013980}

\bibitem[{{Sitnov} {et~al.}(2019){Sitnov}, {Birn}, {Ferdousi}, {Gordeev},
  {Khotyaintsev}, {Merkin}, {Motoba}, {Otto}, {Panov}, {Pritchett}, {Pucci},
  {Raeder}, {Runov}, {Sergeev}, {Velli}, \& {Zhou}}]{Sitnov19}
{Sitnov}, M.~I., {Birn}, J., {Ferdousi}, B., {et~al.} 2019, \ssr, 215, 31,
  \dodoi{10.1007/s11214-019-0599-5}

\bibitem[{{Snekvik} {et~al.}(2012){Snekvik}, {Tanskanen}, {{\O}stgaard},
  {Juusola}, {Laundal}, {Gordeev}, \& {Borg}}]{Snekvik12}
{Snekvik}, K., {Tanskanen}, E., {{\O}stgaard}, N., {et~al.} 2012, \jgr, 117,
  2219, \dodoi{10.1029/2011JA017040}

\bibitem[{{Steinhauer} {et~al.}(2008){Steinhauer}, {McCarthy}, \&
  {Whipple}}]{Steinhauer08}
{Steinhauer}, L.~C., {McCarthy}, M.~P., \& {Whipple}, E.~C. 2008, \jgr, 113,
  4207, \dodoi{10.1029/2007JA012578}

\bibitem[{{Steinolfson} {et~al.}(1982){Steinolfson}, {Suess}, \&
  {Wu}}]{Steinolfson82}
{Steinolfson}, R.~S., {Suess}, S.~T., \& {Wu}, S.~T. 1982, \apj, 255, 730,
  \dodoi{10.1086/159872}

\bibitem[{Sun {et~al.}(2017)Sun, Fu, Wei, Yao, Rong, Zhou, Slavin, Wan, Zong,
  Pu, Shi, \& Shen}]{Sun17:cs_pressure}
Sun, W.~J., Fu, S.~Y., Wei, Y., {et~al.} 2017, \jgr, 122, 12,212,
  \dodoi{10.1002/2017JA024603}

\bibitem[{{Syrovatskii}(1981)}]{Syrovatskii81}
{Syrovatskii}, S.~I. 1981, Annual review of astronomy and astrophysics, 19,
  163, \dodoi{10.1146/annurev.aa.19.090181.001115}

\bibitem[{{Terasawa} {et~al.}(2000){Terasawa}, {Shibata}, \&
  {Scholer}}]{Terasawa00:AdSR}
{Terasawa}, T., {Shibata}, K., \& {Scholer}, M. 2000, Advances in Space
  Research, 26, 573, \dodoi{10.1016/S0273-1177(99)01087-X}

\bibitem[{{Verneta} {et~al.}(1994){Verneta}, {Antonucci}, \&
  {Marocchi}}]{Verneta94}
{Verneta}, A.~I., {Antonucci}, E., \& {Marocchi}, D. 1994, \ssr, 70, 299,
  \dodoi{10.1007/BF00777884}

\bibitem[{{Wang} \& {Bhattacharjee}(1999)}]{Wang&Bhattacharjee99}
{Wang}, X., \& {Bhattacharjee}, A. 1999, \jgr, 104, 7045,
  \dodoi{10.1029/1998JA900124}

\bibitem[{{Washimi} {et~al.}(1987){Washimi}, {Yoshino}, \& {Ogino}}]{Washimi87}
{Washimi}, H., {Yoshino}, Y., \& {Ogino}, T. 1987, \grl, 14, 487,
  \dodoi{10.1029/GL014i005p00487}

\bibitem[{{Wiegelmann} \& {Schindler}(1995)}]{Wiegelmann&Schindler95}
{Wiegelmann}, T., \& {Schindler}, K. 1995, \grl, 22, 2057,
  \dodoi{10.1029/95GL01980}

\bibitem[{{Woo}(1997)}]{Woo97}
{Woo}, R. 1997, \grl, 24, 97, \dodoi{10.1029/96GL03479}

\bibitem[{{Wu} \& {Guo}(1997{\natexlab{a}})}]{Wu&Guo97}
{Wu}, S.~T., \& {Guo}, W.~P. 1997{\natexlab{a}}, Advances in Space Research,
  20, 2313, \dodoi{10.1016/S0273-1177(97)00902-2}

\bibitem[{{Wu} \& {Guo}(1997{\natexlab{b}})}]{Wu&Guo97:AGU}
---. 1997{\natexlab{b}}, Washington DC American Geophysical Union Geophysical
  Monograph Series, 99, 83, \dodoi{10.1029/GM099p0083}

\bibitem[{{Yeh}(1984)}]{Yeh84}
{Yeh}, T. 1984, \apss, 98, 353, \dodoi{10.1007/BF00651414}

\bibitem[{{Yeh} \& {Pneuman}(1977)}]{Yeh&Pneuman77}
{Yeh}, T., \& {Pneuman}, G.~W. 1977, \solphys, 54, 419,
  \dodoi{10.1007/BF00159933}

\bibitem[{{Yin} \& {Winske}(2002)}]{Yin&Winske02}
{Yin}, L., \& {Winske}, D. 2002, Journal of Geophysical Research (Space
  Physics), 107, 1485, \dodoi{10.1029/2002JA009507}

\bibitem[{{Yushkov} {et~al.}(2021){Yushkov}, {Petrukovich}, {Artemyev}, \&
  {Nakamura}}]{Yushkov21}
{Yushkov}, E., {Petrukovich}, A., {Artemyev}, A., \& {Nakamura}, R. 2021,
  Journal of Geophysical Research: Space Physics, 126, e2020JA028969,
  \dodoi{https://doi.org/10.1029/2020JA028969}

\bibitem[{{Zaharia} {et~al.}(2005){Zaharia}, {Birn}, \& {Cheng}}]{Zaharia05}
{Zaharia}, S., {Birn}, J., \& {Cheng}, C.~Z. 2005, \jgr, 110, A09228,
  \dodoi{10.1029/2005JA011101}

\bibitem[{{Zharkova} {et~al.}(2011){Zharkova}, {Arzner}, {Benz}, {Browning},
  {Dauphin}, {Emslie}, {Fletcher}, {Kontar}, {Mann}, {Onofri}, {Petrosian},
  {Turkmani}, {Vilmer}, \& {Vlahos}}]{Zharkova11SSR}
{Zharkova}, V.~V., {Arzner}, K., {Benz}, A.~O., {et~al.} 2011, \ssr, 159, 357,
  \dodoi{10.1007/s11214-011-9803-y}

\end{thebibliography}

\end{document}